\documentclass[english]{article}
\usepackage{graphicx,color}
\def\no{\noindent}
\def\bc{\begin{center}}
\def\ec{\end{center}}

\def\beq{\begin{equation}}
\def\eeq{\end{equation}}

\def\br{{\bf r}}
\def\bx{{\bf x}}
\def\bR{{\bf R}}
\def\bq{{\bf q}}

\def\be{{\bf e}}
\def\bv{{\bf v}}
\def\bw{{\bf w}}
\def\bu{{\bf u}}
\def\by{{\bf y}}

\begin{document}

\title{Repeated measurements and random scattering in quantum walks
}

\author{K. Ziegler\\
Institut f\"ur Physik, Universit\"at Augsburg\\
D-86135 Augsburg, Germany\\
email: klaus.ziegler@physik.uni-augsburg.de
}

\maketitle

\no
Abstract:

We study the effect of random scattering in quantum walks on a finite graph
and compare it with the effect of repeated measurements. 
To this end, a constructive approach is employed by introducing a localized 
and a delocalized basis for the underlying Hilbert space. This enables us to
design Hamiltonians whose eigenvectors are either localized or delocalized.
By presenting some specific examples we demonstrate that the localization of
eigenvectors restricts the transition probabilities on the graph and leads to
a removal of energy states from the quantum walk
in the monitored evolution. We conclude that repeated measurements
as well as random scattering provide efficient tools for controlling quantum walks.

\section{Introduction}
\label{sect:introduction}

Classical random walks play an important role in the description of the dynamics in statistical 
physics~\cite{PEARSON1905,10.1093/acprof:oso/9780199234868.001.0001}
with applications in many research areas, ranging from biological~\cite{WOS:000257009000001}, 
chemical~\cite{WOS:A1983QE45100005} and 
physical systems~\cite{glimm2012quantum,itzykson_drouffe_1989,spitzer1964principles} 
to the dynamics of financial markets~\cite{2019v}.
The basic idea of a classical random walk~\cite{glimm2012quantum,itzykson_drouffe_1989} is to
consider the transition probability $0\le P_{\br\br'}\le 1$ of a random walker to move from the site $\br'$
to the site $\br$ in a fixed time interval. The available sites form a lattice or a graph with the constraint
$\sum_{\br}P_{\br\br'}=1$ for the transition probabilities, which is called detailed 
balance~\cite{PhysRevLett.102.160602}.
It reflects that the random walker does not get lost or absorbed during the random walk. 
In order to include quantum effects, we replace the transition probabilities by a unitary evolution 
from the local state $|\br'\rangle$ to the local state $|\br\rangle$ during the time $\tau$. 
This can be expressed by the unitary operator $U(\tau)=\exp(-iH\tau)$ with the Hermitian Hamiltonian $H$.
For the elements of a unitary matrix $U_{\br\br'}=\langle\br|\exp(-iH\tau)|\br'\rangle$ we apply the
same constraint $\sum_{\br'}U_{\br\br'}=1$ as for the classical transition probability. Thus, the substitution 
$P_{\br\br'}\to U_{\br\br'}$ constitutes a ``quantization'' of the classical random walk.
Most interesting is the asymptotic behavior of $U(\tau)^n=U(n\tau)$ for large $n$, where the time reads 
$t=n\tau$.
In the following we will consider a continuous time $t$ for simplicity. This is a special model from the more
general field of quantum walks, which has become the subject of intensive research 
recently~\cite{PhysRevA.48.1687,WOS:000289698100015,WOS:000308875100002,WOS:000322605800003,
WOS:000692538100009,WOS:000184453100002,MULKEN201137,Das_2022a}.

Properties of the quantum walk depend on the eigenstates and eigenvalues of the Hamiltonian $H$, which
are also the eigenstates of the evolution operator. An important characterization is how far an initial state 
$|\br_0\rangle$ can be transferred to other local states $|\br\rangle$ by the unitary evolution over a large 
time. In other words, is the quantum walk restricted inside the Hilbert space or not?
This depends on whether the eigenbasis of $U(t)$ is localized or delocalized. In the former case the initial
state will only evolve to states $\{|\br\rangle\}$ in a restricted vicinity of $|\br_0\rangle$, while for a delocalized basis
it can reach any of the existing states.
For a given graph, comprising of points $\{\br_1,\ldots,\br_N\}$ embedded in a $d$--dimensional space, 
there are equivalent basis sets for the corresponding Hilbert space, spanned by $\{|\br_1\rangle,\ldots,|\br_N\rangle\}$, 
which are either localized or delocalized or a mixture of both.    
Here is an example of two equivalent basis sets, one is localized, the other delocalized:
The states $\{|\br_1\rangle,\ldots,|\br_N\rangle\}$ can be represented
by a Cartesian basis $\{\be_k\}$, which are localized vectors.
Alternatively, the plane-wave basis $\{\bw_k\}_{k=1,\ldots,N}$, consisting of extended vectors with
\beq
\label{plane_wave0}
\bw_k:=(1,e^{2\pi i (k-1)/N},e^{2\pi i (k-1)2/N},\ldots , e^{2\pi i (k-1)(N-1)/N})^T/\sqrt{N}
,
\eeq
also spans this Hilbert space.
A Hermitian operator $H$ or a unitary operator $U$ can be expanded in any of these basis sets. 
Then the question is,
in which basis the operator is diagonal? Or more specific, is the diagonal basis localized or not?
In this paper we will reverse the mapping $U(\tau)\to\{\bx_k,E_k\}$
by constructing an operator $U(\tau)$ for a given set of eigenvalues $\{E_k\}$ in a specific basis $\{\bx_k\}$: 
$\{\bx_k,E_k\}\to U(\tau)$. For a given spectrum this enables us to choose a localized, a delocalized or a mixed 
basis to construct the operator $U(\tau)$. In this context we will demonstrate that for degenerate eigenvalues
both types of basis sets give the same operator. This means that the distinction between a localized vs.
a delocalized diagonal basis is not applicable in this case and the mapping $\{\bx_k,E_k\}\to U(\tau)$
is not invertible. This reflects the more general problem that the mapping $U(\tau)\to\{\bx_k,E_k\}$
does not have a unique solution in the case of degenerate eigenvalues.
After setting up the unitary operator for a localized and for a plane-wave basis, we will study the effect of (i) 
repeated projective measurements and (ii) random scattering on the quantum walk.

Before we proceed with a detailed discussion of the quantum walk model we give a short summary of the present work.
After defining a quantum walk model on a finite graph with $N$ sites, we begin with introducing a 
symmetric $N\times N$ Hamiltonian that connects all sites of the graph with equal transition amplitudes. 
This Hamiltonian has one zero eigenvalue and an $N-1$--fold degenerate eigenvalue 1. 
A basis of linear independent eigenvectors is immediately identified and
within the Gram-Schmidt approach we construct a localized orthonormal basis for this Hamiltonian. 
On the other hand, the plane-wave basis, representing delocalized eigenvectors, is also an eigenbasis 
for this Hamiltonian. This reflects the fact that this special Hamiltonian does not have a unique eigenbasis
due to the $N-1$--fold degeneracy. In particular, it enables us to choose either a localized or a delocalized
eigenvector. Now we reverse the procedure of finding eigenvectors and eigenvalues for a given
Hamiltonian by constructing a new Hamiltonian for a given basis and given eigenvalues:
Using the two types of eigenvectors, we replace the
degenerate by non-degenerate eigenvalues. In other words, we employ a mapping, either for the 
localized or for the delocalized basis, and a set of arbitrary eigenvalues to create a new 
Hamiltonian $H$. 
Choosing random eigenvalues, we study in Sect.\ref{sect:random} the transition probability
between different states under
a unitary evolution and average this quantity. It turns out that the evolution splits into two parts,
where one is classical with a finite limit for infinite times, while the other describes quantum
fluctuations. The latter decays for longer times, such that only the classical contribution survives.
This is the case for a localized as well as for a delocalized basis. Then the unitary evolution is
compared with a monitored evolution, where repeated projective measurements are applied.
In analogy to the unitary evolution, the monitored evolution is defined by a non-unitary evolution
operator. It is challenging to ask whether the projective measurements play a similar role as the 
random eigenvalues in the unitary evolution. To answer this question, we study in Sect.\ref{sect:mon_ev}
the monitored evolution in terms of the first detected transition from the initial state to the state of 
the measurement.
We find that the localized basis implies a localization effect for the monitored evolution,
while the plane-wave basis provides a uniform spreading of the quantum walk. The behavior
of the classical part during the averaged unitary evolution also distinguishes between a localized and
the plane-wave basis.

The article is organized as follows. In Sect.\ref{sect:quwalk} we introduce the quantum walk
on a finite graph, define the unitary and the monitored evolution, and construct a localized and 
a delocalized basis on the graph. Next we discuss briefly a mapping from eigenvectors and 
non-degenerate eigenvalues to the unitary evolution matrix in Sect.\ref{sect:non-degen}.
This enables us to describe in Sect.\ref{sect:mon_ev} the monitored evolution of the transition probability
for both types of eigenvectors and in Sect.\ref{sect:random} the average transition probability for
the unitary evolution. The results of our calculations are discussed in Sect.\ref{sect:discussion}
and examples of uncorrelated and correlated random eigenvalues are presented in Sect.\ref{sect:examples}.
Finally, in Sect.\ref{sect:conclusions} we conclude our results for the quantum walk and give a brief
outlook for the extension of this work.

\section{Quantum walks on a graph}
\label{sect:quwalk}

The general concept of a quantum walk is quite broad and has many models and
realizations~\cite{PhysRevA.48.1687,PhysRevLett.102.180501,
WOS:000184453100002,MULKEN201137,Das_2022a}.
We will focus here on the tight-binding model, including random 
scattering~\cite{WOS:000302214100002,PhysRevResearch.5.023150}
and repeated projective 
measurements~\cite{Gruenbaum2013,Dhar_2015,Friedman_2016,PhysRevResearch.4.023129,
PhysRevA.110.022208}.
To this end, we define a graph ${\cal G}$ as a set of spatial sites, embedded in a $d$--dimensional space, as
\beq
{\cal G}:=\{\br_1,\br_2,\ldots,\br_N\}
\ \ \ (N<\infty)
,
\eeq
where $\br_j$ is a $d$--dimensional vector. 
Then the quantum walk on ${\cal G}$ is defined by a unitary evolution operator $\exp(-iHt)$ 
with the Hamiltonian $H$, which is acting on the $N$--dimensional Hilbert space $\mathcal{H}$
that is spanned by the basis of the position states $\{|\br_n\rangle\}_{n=1,2,\ldots,N}$.
It describes quantum tunneling between the sites of the graph. ${\cal H}$ is also spanned
by the eigenstates $\{|E_k\rangle\}$ of $H$, whose corresponding eigenvalues are $\{E_k\}$.
Then we write for the unitary evolution matrix
\beq
\label{ue1}
U_{MM'}(t)=\langle\br_M|e^{-iHt}|\br_{M'}\rangle
=\sum_{k=1}^Nq_{M,k}e^{-iE_kt}q^*_{M',k}
,
\eeq
where $q_{M,k}=\langle \br_M|E_k\rangle=\langle E_k|\br_M\rangle^*$ 
is the overlap or scalar product of the particle position state and the energy eigenstate.

\begin{figure}[t]
\centering
\includegraphics[width=8cm,height=3.5cm]{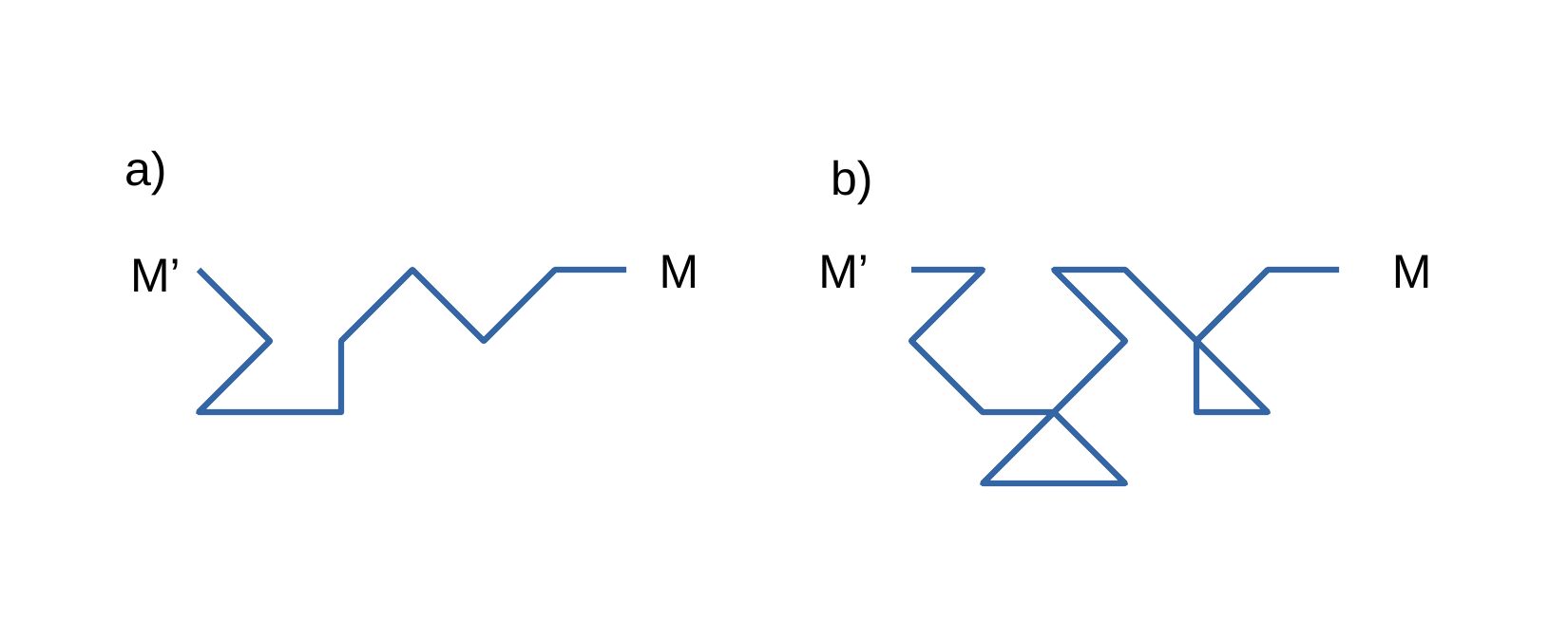}
    \caption{Two typical examples of a monitored quantum walk from $|\br_{M\rq{}}\rangle$ to
    $|\br_M\rangle$ with a) $6$ and b) $14$ steps 
    between projective measurement, where the links represent a unitary evolution for the time $\tau$. 
    The measured state $|\br_M\rangle$ is only visited at the end of the quantum walk. 
    }    
\label{fig:qw}
\end{figure}

Complementary to the unitary evolution we consider a monitored evolution with repeated projective measurements. The idea is to prepare the 
quantum system in an initial state $|\br_{M\rq{}}\rangle$, let it evolve unitarily for the time $\tau$ to 
the state $e^{-iH\tau}|\br_{M\rq{}}\rangle$ and perform a projective measurement with the projector 
$\Pi={\bf 1}-|\br_M\rangle\langle\br_M|$, where ${\bf 1}$ is the identity operator and
$|\br_M\rangle$ a state that defines the measurement. This operation yields the state 
$|\psi_1'\rangle=\Pi e^{-iH\tau}|\br_{M\rq{}}\rangle$, which is either orthogonal to 
$|\br_M\rangle$ or it vanishes when $e^{-iH\tau}|\br_{M\rq{}}\rangle=e^{i\varphi}|\br_M\rangle$ 
with some phase $\varphi$. A further unitary evolution for the time $\tau$ yields $|\psi_1\rangle
=e^{-iH\tau}\Pi e^{-iH\tau}|\br_{M\rq{}}\rangle$ and $\phi_1=\langle\br_M|\psi_1\rangle$. 
If $\phi_1\ne 0$ the system was not in the state $e^{i\varphi}|\br_M\rangle$ when the projection was
applied. This means that our measurement to detect $|\br_M\rangle$
was not successful. In this case we apply another projection to $|\psi_1\rangle$, followed by
a unitary evolution to get $|\psi_2\rangle=e^{-iH\tau}\Pi|\psi_1\rangle$ and 
$\phi_2=\langle\br_M|\psi_2\rangle$.
Again, if $\phi_2\ne0$ the system was not detected in the state $e^{i\varphi}|\br_M\rangle$. These
steps can be repeated $m$ times until the measurement has detected the state $|\br_M\rangle$
such that $\phi_k=0$ for all $k\ge m$. In other words, if the measurement is unsuccessful by not
detecting the state $|\br_M\rangle$, the experiment continues by another measurement, followed by 
the evolution for the time step $\tau$. 
This enables us to determine the case when {\it none} of the first $m-1$ measurements did detect the 
state $|\br_M\rangle$
when $\phi_{m-1}\ne 0$. Moreover, the values of $|\phi_{m_1}|^2$ gives the probability that the system
is in the state $|\br_M\rangle$ after $m-1$ unsuccessful measurements.  The monitored transition 
amplitude can be defined as \cite{PhysRevA.110.022208}
\beq
\label{trans_amp00}
\phi_{MM\rq{}}(m,\tau)
=\langle\br_M|e^{-iH\tau/2}T_M^{m-1} e^{-iH\tau/2}|\br_{M\rq{}}\rangle
\ ,\ \
T_M:=e^{-iH\tau/2}({\bf 1}-|\br_M\rangle\langle\br_M|) e^{-iH\tau/2}
,
\eeq
where $T_M$ is the monitored evolution operator, the analogue to the unitary evolution operator
$e^{-iH\tau}$. This evolution generates a distribution of finite quantum walks with
respect to the number of unsuccessful measurements $m-1$, as illustrated by two examples in 
Fig.\ref{fig:qw}.
In contrast to a unitary evolution at times $t_m=\tau m$, the measured states $|\br_M\rangle$ can only
be reached at the end of the monitored quantum walk.
This protocol was discussed in Ref.~\cite{Gruenbaum2013} and has been applied to single-particle states
to detect the particle location on a 
graph~\cite{Dhar_2015,dhar15,Friedman_2016,lahiri19,PhysRevResearch.1.033086,
PhysRevResearch.2.033113,PhysRevResearch.5.023141}.
Its advantage, in comparison to a unitary evolution, is that 
all information about the transition $|\br_{M\rq{}}\rangle\to|\br_M\rangle$
is collected in the distribution with respect to the number of measurements.

The non-unitary evolution operator $T_M$ has matrix elements in the 
energy basis~\cite{PhysRevA.110.022208}
\beq
\label{me1}
{\hat T}_{M;kl}(\tau):=
\langle E_{k}|e^{-iH\tau}|E_{l}\rangle -\langle E_{k}|e^{-iH\tau/2}|\br_M\rangle
\langle \br_M|e^{-iH\tau/2}|E_{l}\rangle
=e^{-iE_M\tau/2}(\delta_{kl}-q^*_{M,k}q_{M,l})e^{-iE_{l}\tau/2}
.
\eeq
This provides the $N\times N$ matrix
${\hat T}_M(\tau)={\hat U}(\tau/2)({\bf 1}-Q_M^*EQ_M){\hat U}(\tau/2)$,
where ${\bf 1}$ is the $N\times N$ unit matrix and $E$ is the $N\times N$ matrix whose matrix elements are 1,
and with the diagonal matrices
\[
{\hat U}(\tau)={\rm diag}(e^{-iE_1\tau},\ldots,e^{-iE_N\tau})
\ ,\ \ 
Q_M={\rm diag}(q_{M,1},\ldots, q_{M,N})
.
\]
It describes a unitary evolution for the time $\tau/2$,
followed by a projection onto the space orthogonal to $|\br_M\rangle$ by ${\bf 1}-Q_M^*EQ_M$, 
and finally a unitary evolution for the time $\tau/2$ again.

Both evolution matrices $U$ and $T$ are defined via the eigenvalues $\{E_k\}$ of the Hamiltonian $H$ and the 
overlap functions 
$\{ q_{n,k}\}$ between the position states and the eigenstates of the Hamiltonian. The eigenvalues of $U(t)$ are 
on the unit circle and the eigenvalues of $T_M$ are on the unit disk with one eigenvalue at the center, since
$ {\bf 1}-Q_M^*EQ_M$ has an eigenvalue 0 with the eigenvector 
${\bq}^*_M=(q^*_{M,1},\ldots, q^*_{M,N})^T$ 
and the $N-1$--fold degenerate eigenvalue 1 with eigenvectors $\{\bq^*_{k}\}_{k\ne M}$.

To define a quantum walk on ${\cal G}$ 
we use the graph's adjacency matrix with specific weights for its elements.
We begin with the idealistic case, in which the diagonal 
elements of $H$ are $1-1/N$ and the off-diagonal elements connect all sites equally with weight 1:
\beq
\label{ideal_ham}
H_{\br_j\br_{j'}}\equiv H_{jj'}=\delta_{jj'}-\frac{1}{N}
.
\eeq
This can also be written as the symmetric $N\times N$ matrix
$
H={\bf 1}-E/N
$.
The vector $\bu_1:=(1,\ldots ,1)^T$ is eigenvector with eigenvalue 0: $H\bu_1=0$.
This reflects the detailed balance property $\sum_{j'}U_{jj'}=1$ of the quantum walk at each site of 
the graph, which restricts the quantum walk effectively to an $N-1$ dimensional Hilbert space.

Besides the zero eigenvalue of $H$ with eigenvector $\bu_1$, there is the $N-1$--fold degenerate eigenvalue 1.
Thus, we must construct a basis of linear independent vectors, which are orthogonal to $\bu_1$. 
An example is $\{\bu_l\}_{l=2,\ldots ,N}$ with $\bu_l:=(1,0,\ldots,0,-1,0\ldots 0)^T$, where the $l^{\rm th}$ 
component is $-1$. This means that the Hilbert space ${\cal H}$ separates into $\mathcal{H}_1$, spanned by $\bu_1$ 
and $\mathcal{H}'$, spanned by $\{\bu_l\}_{l=2,\ldots ,N}$, as $\mathcal{H}=\mathcal{H}_1\oplus\mathcal{H}'$.
The basis $\{\bu_l\}_{l=2,\ldots ,N}$ is not orthogonal, though, but can be orthonormalized by the 
Gram-Schmidt approach.
This recursive procedure creates $N$-component vectors that are systematically filled with non-zero
components, beginning, for instance, with $\bv_2=\frac{1}{\sqrt{2}}(1,-1,0,\ldots ,0)^T$ and proceeding with
\beq
\label{ONS0}
\bv_k=\frac{1}{\sqrt{k(k-1)}}(1,\ldots,1,-k+1,0,\ldots,0)^T
\ \ \ (3\le k\le N)
,
\eeq
where the first $k-1$ components are $1$. 
This enables us to expand the adjacency matrix (or Hamiltonian)
$(H_{jj'})$ of Eq. (\ref{ideal_ham}) in terms of its eigenvalues and eigenvectors as
\beq
\label{Ham_exp0}
H_{jj'}
=\sum_{k=2}^Nv^*_{k,j}E_{k}v_{k,j'}
,\ \
E_1=0
\ ,\ \
E_k=1
\ \ \
(2\le k\le N)
,
\eeq
where the first index $k$ of $v_{k,l}$ refers to the vector $\bv_k$ and the second index $l$ refers to its components.
It should be noted that $v_{k,l}=q^*_{l,k}=\langle \br_l|E_k\rangle^*=\langle E_k|\br_l\rangle$.
Moreover, the matrix elements $v_{1,l}=1/\sqrt{N}$ do not appear in the expansion of $H$ due to $E_1=0$,
which is required by detailed balance.
For $2\le k\le N$, $1\le l\le N$ we get from Eq. (\ref{ONS0})
\beq
\label{unit_el0}
v_{k,l}=\frac{1}{\sqrt{k(k-1)}}\cases{
1 & $1\le l\le k-1$ \cr
-k+1 & $l=k$ \cr
0 & $k< l\le N$ \cr
}
.
\eeq
This can also be written as
\beq
\label{unit_el1a}
v_{k,l}=\frac{1}{\sqrt{k(k-1)}}\left[(-k+1)\delta_{kl}+\Theta_{kl}\right]
\ \ {\rm with}\ \ \Theta_{kl}=\cases{
0 & $l\ge k$ \cr
1 & $l< k$ \cr
} 
.
\eeq
Although we don't need $\bv_{1}=(1,\ldots,1)^T/\sqrt{N}$ for the expansion of $H$, it contributes
to the expansion of the unitary evolution matrix $U(t)=\exp(-iHt)$.
After including $k=1$ in the matrix $v$ we obtain from Eq. (\ref{unit_el1a}) for its elements
\beq
\label{unit_el2}
v_{k,l}=\frac{1}{\sqrt{N}}\delta_{k1}
+(1-\delta_{k1})\frac{1}{\sqrt{k(k-1)}}\left[(-k+1)\delta_{kl}+\Theta_{kl}\right]
.
\eeq
While the Gram-Schmidt procedure gives a unique basis (which depends on the choice of $\bu_2$
though), another basis can be created by a unitary transformation. An important question is
related to the localization of eigenvectors. $\{\bu_l\}_{l=2,\ldots ,N}$ are localized eigenvectors
but summing them creates also extended eigenvectors, as already demonstrated for the orthonormal basis
$\{\bv_l\}_{l=2,\ldots ,N}$. This ambiguity exists only for degenerate eigenvalues, where any
linear combination of eigenvectors gives another eigenvector.

For specific values of $k$ and $l$ the spectral weights $(v_{k,l})^2=\langle\br_k|E_l\rangle\langle E_k|\br_l\rangle$ 
characterizes the decay of the eigenvector $\bv_k$.
For instance, $\br_1$ is equally connected through $U(t)$ to any site $\br_l$ on the graph, while $\br_2$ is 
only connected to
$\br_1$, $\br_3$ to $\br_1$ and $\br_2$ etc. up to $\br_N$ which is connected to all sites but also has
a dominant spectral weight $(v_{N,N})^2=1-1/N$. This, as well as $(v_{k,k})^2=1-1/k$ ($2\le k\le N$),
reflects localization with increasing spectral weight due to increasing $k$. More general, the quantity
\beq
\label{loc_criterion}
c_k:=\sum_{l=1}^N(v_{k,l})^4
\eeq
provides a criterion for the localization of $\bv_k$: It defines localization for $c_k\sim 1/N^0$ 
and delocalization for $c_k\sim 1/N$. For the basis $\{\bv_k\}$ we get $c_1=1/N$ and 
for $2\le k\le N$
\beq
\label{loc_criterion1}
c_k=\frac{1}{k^2(k-1)^2}\sum_{l=1}^N[(k-1)^4\delta_{kl}+\Theta_{kl}]
=(1-1/k)^2+\frac{1}{k^2(k-1)}
,
\eeq
which is independent of $N$ and bounded as $1/2\le c_k\le 1$.
Thus, all eigenvectors are localized except for $\bv_1$. The localization originates in the
component $v_{k,k}$ of the degenerate spectrum, which is larger than the other components of $\bv_k$.
In contrast to the localized $\{\bv_k\}_{l=2,\ldots,N}$, the basis $\{\bw_l\}$ is a set of delocalized eigenvectors 
with $c_k=\sum_l|w_{kl}|^4=1/N$.

${\cal H}$ can also be spanned by plane waves $\{\bw_k\}$ in Eq. (\ref{plane_wave0}),
which form an orthonormal basis.
This provides the elements $w_{jk}:=\exp[2\pi i(j-1)(k-1)/N]/\sqrt{N}$ ($1\le j,k\le N$) of
the $N\times N$ unitary matrix
\[
w=\frac{1}{\sqrt{N}}\pmatrix{
1 & 1 & 1 & \cdots & 1 \cr
1 & e^{2\pi i/N} & e^{2\pi i2/N} & \ldots & e^{2\pi i(N-1)/N} \cr
\vdots & \vdots & \vdots & \ddots & \vdots \cr
1 & e^{2\pi i(N-1)/N} & e^{2\pi i2(N-1)/N} & \ldots & e^{2\pi i(N-1)^2/N} \cr
}
\]
with
\[
\sum_{k=1}^{N}w_{jk}w_{j'k}^*=\sum_{k=1}^{N}\frac{e^{2\pi i(j-j')(k-1)/N}}{N}=\delta_{jj'}
.
\]
This enables us to expand the unitary evolution matrix as
\beq
\label{Ham_exp1}
U_{jj'}=\sum_{k=1}^{N}w_{kj}^*e^{-iE_k\tau} w_{kj'}
,
\eeq
which gives the same Hamiltonian as the expansion in Eq. (\ref{Ham_exp0}) due to the spectral
degeneracy in ${\cal H}'$.

\subsection{Non-degenerate energy eigenvalues}
\label{sect:non-degen}

Based on the expansions (\ref{Ham_exp0}) and (\ref{Ham_exp1}) with a non-degenerate set of eigenvalues 
$\{E_j\}$ we create new Hamiltonians, only keeping $E_1=0$ to preserve the detailed balance,
either with the localized eigenvector basis $\{\bv_k\}$ or with the delocalized eigenvector basis $\{\bw_k\}$:
\beq
\label{Ham_exp2}
U'_{jj'}=\sum_{k=1}^{N}v_{k,j}e^{-iE_k\tau} v_{k,j'}
\ ,\ \ \
U''_{jj'}=\sum_{k=1}^{N}w_{kj}^*e^{-iE_k\tau} w_{kj'}
,
\eeq
where $U''$ is translational invariant on the graph in the sense of $U''_{jj'}\equiv H''_{j-j'}$.
(It should be noted that this does not mean translational invariance with respect the embedding space.)
Formally, we can also consider $E_1\ne0$. In that case the Hamiltonian also depends on the 
delocalized eigenvector $\bv_1$. 

The expansions in Eq. (\ref{Ham_exp2}) mean that we
can create two different Hamiltonians with the same spectrum but different properties of the
eigenvectors; in one case with a delocalized basis $\{\bw_l\}$, in the other case with a localized basis $\{\bv_l\}$.
This concept can be generalized by mixing the two type of basis states in different subspaces
of ${\cal H}'$ to create a random ensemble, where the average can either be dominated by the localized
or by the delocalized eigenvectors. Formally, we write $\mathcal{H}'=\mathcal{H}_2\oplus\cdots\oplus\mathcal{H}_n$
and choose randomly either a localized or delocalized basis to span $\mathcal{H}_j$. Finally, we
employ the Gram-Schmidt procedure to orthormalize the basis in $\mathcal{H}'$.


\subsection{Transition probabilities for the monitored evolution}
\label{sect:mon_ev}

From the unitary evolution matrix of Eq. (\ref{ue1}) and the monitored evolution matrix in Eq. (\ref{me1})
we obtain the corresponding transition probabilities as
\beq
\label{unit_tr1}
P_{MM'}(t)=|U_{MM'}(t)|^2=\left|\sum_{k=1}^Nq_{M,k}e^{-iE_kt}q^*_{M',k}\right|^2
\eeq
and, according to Ref. \cite{PhysRevA.110.022208}, we get from Eq. (\ref{trans_amp00})
\beq
\label{mon_tr1}
\Pi_{MM'}(m,\tau)=\left|{\rm Tr}({\hat U}(\tau/2){\hat T}_M^{m-1}{\hat U}(\tau/2)
Q^*_{M'}EQ_M)\right|^2
=\left|\sum_{k,l=1}^N e^{-i(E_k+E_l)\tau/2}[{\hat T}_M^{m-1}]_{kl}q^*_{M',l}q_{M,k}\right|^2
.
\eeq
The conservation of the quantum walker requires $\sum_MP_{MM'}(t)=1$, which is automatically implied by the unitary 
evolution. This condition reflects that the quantum walker is not absorbed at any time and that it can
be found somewhere on the graph with total probability 1. This corresponds with the detailed balance of a classical
random walk, while the detailed balance of the quantum walk $\sum_{j'}U_{jj'}(t)=1$ is an additional
condition.

The time of the monitored evolution is $t=m\tau$, which is interrupted by $m-1$ measurements.
The two transition probabilities are related as $\Pi_{MM'}(1,\tau)=P_{MM'}(\tau)$, which reflects
that $m=1$ means no measurement.
For the monitored transition $|\br_{M'}\rangle\to|\br_M\rangle$ the matrix ${\hat T}_M^{m-1}$ is crucial.
Its matrix elements $q^*_{M',k}q_{M,l}$ read in the plane-wave basis $\{\bw_l\}$
\beq
\label{transition_matrix1}
q^*_{M',k}q_{M,l}=w_{kM'}w^*_{lM}=\frac{e^{2\pi i[(M'-1)(k-1)-(M-1)(l-1)]/N}}{N}
,
\eeq
which implies that the transition probability reads
\beq
\label{transition_matrix1}
\Pi_{MM'}(m,\tau)
=\frac{1}{N^2}\left|\sum_{k,l=1}^N e^{-i(E_k+E_l)\tau/2+2\pi i[(M'-1)(k-1)-(M-1)(l-1)]/N}
[{\hat T}_M^{m-1}]_{kl}\right|^2
.
\eeq
For the corresponding expression in the localized basis $\{\bv_l\}$ we use Eq. (\ref{unit_el2}).
The decay of the monitored transfer probability  with respect to $|M-M'|$ is determined by the matrix 
${\hat T}_M^{m-1}$ in Eq. (\ref{mon_tr1}).
When $m\ll N$ we can neglect terms of $O(1/\sqrt{N})$. Then the transfer matrix 
${\hat T}_M$ is reduced to the matrix $(v_{k,l})$ for ($2\le k,l\le N$) as
\[
A_{kl}:=\delta_{kl}-q^*_{M,k}q^{}_{M,l}=\delta_{kl}-v_{k,M}v_{l,M}
\]
\[
=\delta_{kl}[1-(1-1/M)\delta_{kM}]
+\frac{\sqrt{1-1/M}}{\sqrt{k(k-1)}}\Theta_{kM}\delta_{lM}
+\frac{\sqrt{1-1/M}}{\sqrt{l(l-1)}}\Theta_{lM}\delta_{kM}
-\frac{1}{\sqrt{k(k-1)l(l-1)}}\Theta_{kM}\Theta_{lM}
.
\]
Thus, the $(N-1)\times(N-1)$ matrix $A$ can be written as a matrix with a block structure as
\beq
\label{loc_mon}
A=\pmatrix{
{\bf 1} & 0 \cr
0 & B \cr
}
\ \ {\rm and}\ \ 
{\hat T}_M=U(\tau/2)AU(\tau/2)
=\pmatrix{
u_1 & 0 \cr
0 & u_2Bu_2 \cr
}
\eeq
with an $M\times M$ block matrix $B$ and diagonal matrices $u_{1,2}$ with
$u_1={\rm diag}(e^{-iE_1\tau},\ldots,e^{-iE_{M-1}\tau},0,\ldots)$ and
$u_2={\rm diag}(0,\ldots,0,e^{-iE_M\tau/2},\ldots,e^{-iE_N\tau/2})$.
This means that a vector $\bx=\sum_{k=2}^{M-1} x_k\bv_k$ is completely localized on the graph
under a monitored evolution: ${\hat T}_M\bx=u_1\bx$,
and that the energy states $\{|E_k\rangle\}_{k=2,\ldots,M-1}$ are removed
from the quantum walk.
On the other hand, the complementary vector $\by=\sum_{k=M}^N y_k\bv_k$
can not leave the subspace spanned by $\{\bv_k\}_{k=M,\ldots,N}$ under the monitored evolution:
$\bv_k\cdot {\hat T}_M^{m-1}\by=0$ for $k=2,\ldots,M-1$ and any $m>1$.
For the plane-wave basis $\{\bw_l\}$ the corresponding matrix reads
\[
A_{kl}=\delta_{kl}-q^*_{M,k}q^{}_{M,l}=\delta_{kl}-w^*_{kM}w^{}_{lM}
=\delta_{kl}-\frac{1}{N}e^{-2\pi i (k-l)(M-1)/N}
,
\]
such that
\beq
\label{deloc_mon}
{\hat T}_{M;kl}=e^{-i{\bar E}_k\tau}\delta_{kl}-\frac{1}{N}e^{-2\pi i (k-l)(M-1)/N-i({\bar E}_k
-{\bar E}_l)\tau/2}
.
\eeq
In this case we get $\bv_k\cdot {\hat T}_M^{m-1}\by\ne 0$, which indicates that there is no restriction 
of the monitored evolution.


\subsection{Random scattering in the unitary evolution}
\label{sect:random}

The unitary evolution is subject to strong quantum fluctuations due to the phase factors $\exp(-iE_kt)$.
This can cause a problem when we are interested in the characterization of generic properties. 
A typical example is the entanglement entropy, which can be a wildly fluctuating quantity. In particular,
these fluctuations are very strong near phase transitions.
They are substantially reduced when we average with respect to a time interval $\Delta t$, which
enabled us to identify a sharp phase transition to Hilbert-space
localization~\cite{sym13101796,PhysRevA.107.012413}. Alternatively to a random time interval
we will consider in this section
random scattering in the unitary evolution by random eigenvalues $\{E_k\}$.
In general, both matrices ${\hat E}={\rm diag}(E_1,\ldots,E_N)$ 
and $u$ in $H=u^\dagger {\hat E}u$ can be random, where only ${\hat E}$ appears in the
time dependence of the evolution matrix of Eq. (\ref{ue1}), though. 
Thus, we get from Eq. (\ref{unit_tr1}) the average with respect to the eigenvalue distribution  
for a general unitary matrix $u$
\beq
\label{tr_prob02a}
\langle P_{MM'}(t)\rangle
=\langle [e^{-iHt}]_{MM'}[e^{iHt}]_{M'M}\rangle
=\sum_{k,l}p_{kl}(t)q^{}_{M,k}q^*_{M',k}q^*_{M,l}q^{}_{M\rq{},l}
\eeq
with $p_{kl}(t)=\langle e^{-i(E_k-E_l)t}\rangle
=\langle e^{-i(x_k-x_l)t}\rangle e^{-i({\bar E}_k-{\bar E}_l)t}$ and with
random fluctuations $x_k=E_k-{\bar E}_k$ around the mean eigenvalue ${\bar E}_k$.
This is reminiscent of a time average, since the energy levels appear only in the combination
$E_kt$ in Eq. (\ref{unit_tr1}). Therefore, we expect a similar suppression of the quantum fluctuations
as under time average. However, since $\langle e^{-i(E_k-E_l)t}\rangle_t$ is different from
$p_{kl}(t)$, quantum fluctuations survive at least for short times. This can be seen when
we split the double summation as
\beq
\label{tr_prob03}
\langle P_{MM'}(t)\rangle
=\sum_{k} 
|q^{}_{M,k}|^2|q_{M',k}|^2
+\sum_{k,l}(1-\delta_{kl})p_{kl}(t)q^{}_{M,k}q^*_{M',k}q^*_{M,l}q^{}_{M\rq{},l}
,
\eeq
where we have used $p_{kk}(t)=1$.
Thus, only the remaining double sum depends on time through $p_{kl}(t)$. 
These results can be summarized by the expression
\beq
\label{tr_pr_splitting}
\langle P_{MM'}(t)\rangle
=[1-w(t)]\sum_{k=1}^N|q_{M',k}|^2|q_{M,k}|^2+w(t){\bar P}_{MM'}(t)
,
\eeq
where ${\bar P}_{MM'}(t)$ is the transition probability of Eq. (\ref{unit_tr1}), in which the 
Hamiltonian $H$ is replaced by its average ${\bar H}=\langle H\rangle$.
The weight function $w(t)=\langle e^{-i(x_k-x_{l})t}\rangle$ is determined by the distribution of the eigenvalues.
We can assume that $\lim_{t\to\infty}w(t)=0$, as will be demonstrated in several examples in Sect. \ref{sect:examples}. 
The splitting of the average transition probability into a static part $\sum_{k=1}^N|u_{kM'}|^2|u_{kM}|^2$
and the evolution part ${\bar P}_{MM'}(t)$ with time dependent weights $1-w(t)$ and $w(t)$, respectively,
indicates a separation of the unitary evolution into a classical and a quantum part.
 
For the sum in Eq. (\ref{tr_pr_splitting})
we get from Eq. (\ref{unit_el2}) with $M\ge M'$ in the localized basis $\{\bv_k\}$
\beq
\label{inf_time1}
\lim_{t\to\infty}\langle P_{MM'}(t)\rangle
=\sum_{k=1}^Nv_{k,M}^2v_{k,M'}^2
=\delta_{MM'}\frac{(M-1)^2-1}{M^2}+\frac{1}{N^2}
+\frac{1}{M^2}+\sum_{k=M+1}^N\frac{1}{k^2(k-1)^2}
.
\eeq
The sum on the right-hand side gives a closed expression in terms of the digamma function~\cite{abramowitz+stegun}.
This result is not translational invariant because it favors transitions to small
values of $M,M'$. Of course, this is a consequence of the chosen basis of localized
eigenvectors, which itself is not translational invariant. But it can be
made translational invariant by summing up a permuted basis on the graph, rather 
than fixing a specific one. This would remove the $M$--dependent terms and replace 
it by an average over all positions on the graph.
  
In contrast to the localized basis, for the plane-wave basis we get in the infinite-time limit a uniform 
transition probability:
\beq
\label{inf_time2}
\lim_{t\to\infty}\langle P_{MM'}(t)\rangle
=\sum_{k=1}^N |w_{kM}|^2|w_{kM'}|^2
=\frac{1}{N}
,
\eeq
indicating a uniform distribution for the quantum walk at large times. This reflects the
result for ${\hat T}_M$ in Eq. (\ref{deloc_mon}).

\section{Discussion}
\label{sect:discussion}

According to Eq. (\ref{Ham_exp2}), the unitary evolution matrix $U(t)$ can be expanded in terms of its
eigenbasis. The properties of this basis determine the range of the evolution, i.e., which sites of the graph
are connected by $U(t)$ during the evolution. This can be realized by designing an appropriate quantum 
circuit, enabling us to study the localized as well as delocalized regimes on a quantum computer.

Averaging of the transition probability over a random distribution of eigenvalues of $U(t)$, as described
in Sect. \ref{sect:random}, reveals two 
separate regimes in the evolution. One is static and classical, while the other represents a quantum walk with the 
average Hamiltonian (cf. Eq. (\ref{tr_pr_splitting})). They appear with time-dependent weights, in which the 
contribution of the classic static part becomes more important with increasing time. It means that the
quantum effect is suppressed in the long-time asymptotic under the average. This classical limit reflects
the distribution of the transition probabilities according to their eigenbasis, as given in Eq. (\ref{inf_time1})
for a localized basis and in Eq. (\ref{inf_time2}) for the plane-wave basis.

Instead of random scattering in the unitary evolution of a quantum walk, we also considered the effect of repeated 
projective measurements on the quantum walk in Sect. \ref{sect:mon_ev}. The result of such a monitored
evolution also depends strongly on the properties of the underlying eigenbasis. The corresponding evolution
operator $T_M$, where $|\br_M\rangle$ is the projected state in the measurement, is even more sensitive to the
localization properties, as reflected by the results in Eqs. (\ref{loc_mon}), (\ref{deloc_mon}).
In particular, the  relation ${\hat T}_M^{m-1}\bx=u_1^{m-1}\bx$ for 
$\bx=\sum_{k=2}^{M-1} x_k\bv_k$ implies
a projection onto a subspace that is defined by the measured state.
This could be used to control and guide quantum walks inside a larger Hilbert space by constructing proper 
Hamiltonians $H$ with the mapping $\{E_k,\bv_k\}\to H$ in a localized basis.

\subsection{Examples for the distribution of random eigenvalues}
\label{sect:examples}

\begin{figure}[t]
\centering
a)
\includegraphics[width=7.5cm,height=7.5cm]{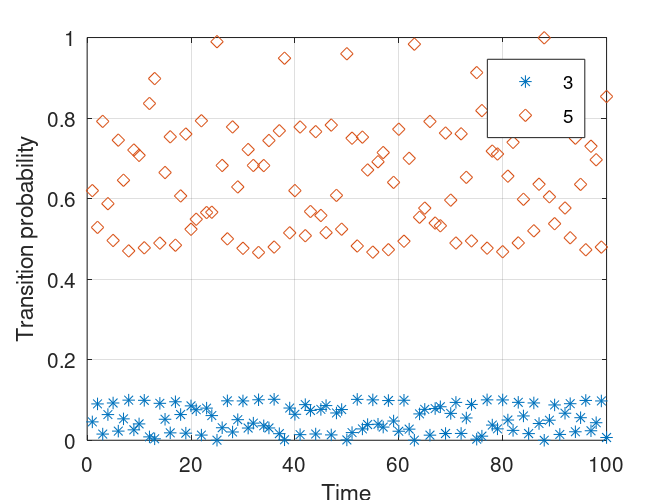}
b)
\includegraphics[width=7.5cm,height=7.5cm]{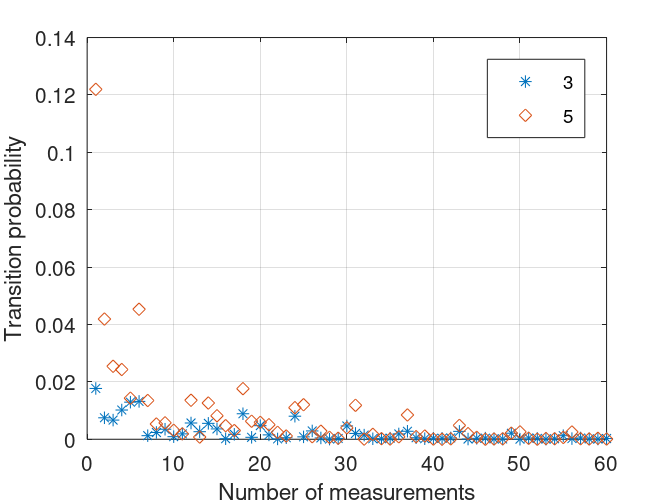}
    \caption{The transition probability for $|\br_j\rangle\to|\br_5\rangle$ ($j=3,5$) and $N=10$
    energy levels in the localized basis for a) the unitary evolution and b) for the monitored evolution.
       }    
\label{fig:unitary}
\end{figure}
\begin{figure}[t]
\centering
a)
\includegraphics[width=7.5cm,height=7.5cm]{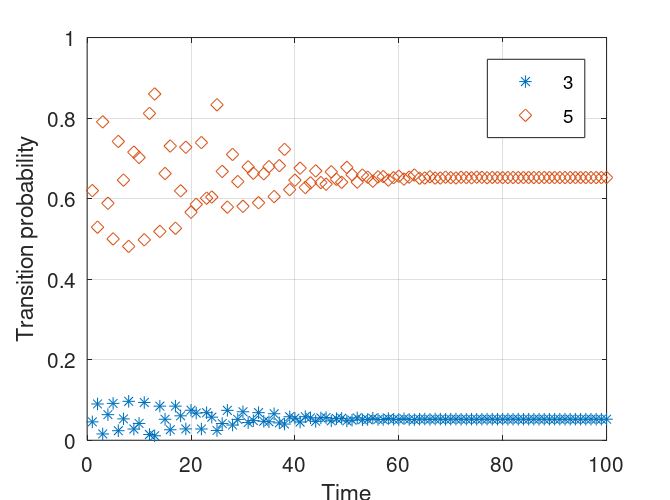}
b)
\includegraphics[width=7.5cm,height=7.5cm]{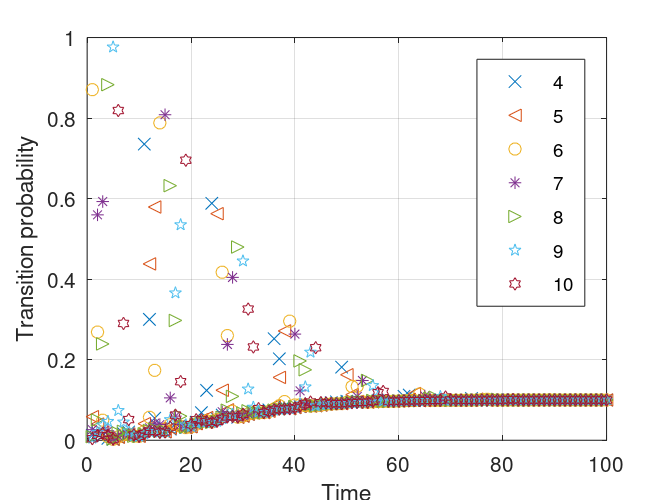}
    \caption{The transition probability of the unitary evolution after averaging with respect to $N=10$ 
    random energy levels.
    a) In the localized basis the transition probability is plotted for $|\br_j\rangle\to|\br_5\rangle$ 
    ($j=3,5$) with two different asymptotic values (cf. Eq. (\ref{inf_time1})).
    b) In the plane-wave basis for the transitions $|\br_j\rangle\to|\br_5\rangle$ ($j=4,\ldots,10$)
    there is only one asymptotic value $0.1$ for all transitions, according to Eq. (\ref{inf_time2}).
    }    
\label{fig:av_unitary}
\end{figure}

We consider the Gaussian distribution of the eigenvalues $\{E_k\}$
\beq
\label{corr_gauss}
\frac{1}{\cal N}\prod_{k,k'} e^{-\gamma_{kk'}
(E_k-{\bar E}_k)(E_{k'}-{\bar E}_{k'})}\prod_{\bar k} dE_{\bar k}
\ \ \ \ \ (-\infty<E_k<\infty)
\eeq
with the positive symmetric correlation matrix $\gamma$ and the normalization ${\cal N}\propto \det\gamma^{-1/2}$.
This gives us
\beq
\label{Gaussian1}
p_{kk'}(t)
=\langle e^{-i(E_k-E_{k'})t}\rangle
=e^{-(\gamma^{-1}_{kk}-\gamma^{-1}_{k'k})t^2/2}e^{-i({\bar E}_k-{\bar E}_{k'})t}
.
\eeq
A constant part of the $k$ dependent average eigenvalues cancels in ${\bar E}_k-{\bar E}_{k'}$
and only the dispersive part contributes. 

The uncorrelated limit is obtained from Eq. (\ref{Gaussian1}) when we assume a diagonal matrix $\gamma$.
Assuming that $\gamma^{-1}_{kk}$ is constant in $k$ with $\gamma^{-1}_{kk}=\kappa$, this provides
\beq
\label{Gaussian2}
p_{kk'}(t)=\cases{
1 & for $k'=k$ \cr
e^{-\kappa t^2/2-i({\bar E}_k-{\bar E}_{k'})t} & for $k'\ne k$ \cr
}
.
\eeq
This means that the function $w(t)$ in the average return probability in Eq. (\ref{tr_pr_splitting})
reads in this case $w(t)=e^{-\kappa t^2/2}$.

A special example for a correlated distribution of the fluctuations $x_k=E_k-{\bar E}_k$ is an attraction 
of eigenvalues with
\beq
\label{attract}
\sum_{k,k'}x_k\gamma_{kk'}x_{k'}
=\frac{1}{\kappa}\left[a\sum_k x_k^2+\frac{1}{2N}\sum_{k,k'}(x_k-x_{k'})^2\right]
\ \ \ (a>0)
,
\eeq
which implies
\[
\gamma_{kk'}=\frac{1}{\kappa}[(1+a)\delta_{kk'}-1/N]
.
\]
Then the elements of the inverse correlation matrix read
\[
\gamma^{-1}_{kk'}=\frac{\kappa}{1+a}\left(\delta_{kk'}+\frac{1}{aN}\right)
\ \ {\rm and}\ 
\gamma^{-1}_{kk}-\gamma^{-1}_{kk'}=\frac{\kappa}{1+a}
\ \ (k'\ne k)
.
\]
Another example is a repulsion of eigenvalues with
\beq
\label{repulse}
\sum_{k,k'}x_k\gamma_{kk'}x_{k'}
=\frac{1}{\kappa}\left[b\sum_k x_k^2-\frac{1}{2N}\sum_{k,k'}(x_k-x_{k'})^2\right]
\ \ (b>1)
,
\eeq
which yields
\[
\gamma_{kk'}=\frac{1}{\kappa}[(b-1)\delta_{kk'}+1/N]
\]
and the elements of the inverse correlation matrix
\[
\gamma^{-1}_{kk'}=\frac{\kappa}{(b-1)}\left(\delta_{kk'}-\frac{1}{Nb}\right)
\ \ {\rm and}\ 
\gamma^{-1}_{kk}-\gamma^{-1}_{kk'}=\frac{\kappa}{b-1}
\ \ (k'\ne k)
.
\]
Then we get from Eq. (\ref{Gaussian1}) for attractive eigenvalues
\beq
\label{attract1}
p_{kk'}(t)
=\cases{
1 & for $k'=k$ \cr
e^{-\kappa t^2/2(1+a)-i({\bar E}_k-{\bar E}_{k'})t} & for $k'\ne k$ \cr
}
\eeq
and for repulsive eigenvalues 
\beq
\label{repulse1}
p_{kk'}(t)
=\cases{
1 & for $k'=k$ \cr
e^{-\kappa t^2/2(b-1)-i({\bar E}_k-{\bar E}_{k'})t} & for $k'\ne k$ \cr 
}
,
\eeq
which is similar to the uncorrelated result in Eq. (\ref{Gaussian2}).
In particular, for all examples the off-diagonal part of $p_{kk'}(t)$ decays with time, 
such that in the end the walker is
distributed over the graph ${\cal G}$ according to Eq. (\ref{inf_time1}) for a localized eigenbasis
or according to Eq. (\ref{inf_time2}) for a plane-wave eigenbasis. For finite values of $t$ there is scaling 
with the variable $\kappa t^2$, where a smaller $\kappa$ means smaller fluctuations of the eigenvalues.
With these results for $p_{kk'}(t)$ we return to the average transition probability of Eq. (\ref{tr_pr_splitting}) and
insert the new weight functions $w(\kappa t^2)$. 
Thus, Gaussian random scattering results in a fast decay of the weight of the quantum
term in $\langle P_{MM'}(t)\rangle$ that is supported by increasing fluctuations.
Therefore, in order to describe a realistic situation, either a finite $t$ or a time average 
by integrating over all times should be considered.
Finally,
in the degenerate case ${\bar E}_k={\bar E}=const.$ we get strict localization of the quantum term in
${\bar P}_{MM'}(t)=\delta_{MM'}$. This indicates that a degenerate average Hamiltonian
reproduces an unphysical result similar to that observed for the Hamiltonian in Eq. (\ref{ideal_ham}).

In order to illustrate the results found in Sects.\ref{sect:mon_ev} and \ref{sect:random},
we consider specific cases for $P_{MM'}(t)$ in Eq. (\ref{unit_tr1}) and for $\Pi_{MM'}(m,\tau)$
in Eq. (\ref{mon_tr1}) with energy levels  $E_k=k-1$ ($1\le k\le 10$) in units of $\hbar$.
In Fig.\ref{fig:unitary}a the unitary evolution of the transition probabilities
for $|\br_3\rangle\to|\br_5\rangle$ and for $|\br_5\rangle\to|\br_5\rangle$ is plotted 
at  discrete times $t=1,2,\ldots, 100$. The probabilities of the same transitions during the
monitored evolution are plotted in Fig.\ref{fig:unitary}b. The decay of the probabilities in the latter case,
in contrast to the oscillating behavior during the unitary evolution, indicates that the
project measurements effectively suppress quantum fluctuations.    

In Fig.\ref{fig:av_unitary} the average transition probability $\langle P_{MM'}(t)\rangle$ of 
Eq. (\ref {tr_pr_splitting}) is plotted during the unitary evolution. The distribution of the random
energy levels uses the examples of Sect.\ref{sect:examples} with ${\bar E}_k=k-1$ ($1\le k\le 10$) 
in units of $\hbar$ and with $w(t)=e^{-t^2/1000}$. The transition probabilities are calculated at discrete 
times $t=1,2,\ldots, 100$, using localized eigenvectors in Fig.\ref{fig:av_unitary}a and plane-wave
eigenvectors in Fig.\ref{fig:av_unitary}b. The two distinct asymptotic values for different initial states
reflects the localization effect in the behavior at large times. For the plane-wave basis, in contrast,
the asymptotic value is universal, as given in Eq. (\ref{inf_time2}).

\section{Conclusions}
\label{sect:conclusions}

We suggest an approach, based on the mapping from a given set of eigenvalues and eigenvectors to the evolution
operator $\{\bx_k,E_k\}\to U(\tau)$, to construct a quantum walk model. This is an alternative to the usual
mapping from a given Hamiltonian to the eigenvalues and eigenvectors $H\to\{\bx_k,E_k\}$. 
It provides a more direct insight into the properties of the
quantum walk and avoids the costly calculation of the eigenvectors and eigenvalues of $H$.
As demonstrated in Sects. \ref{sect:mon_ev} and \ref{sect:random}, this
approach reveals how the localization properties of the unitary and the monitored evolution 
can be controlled by designing the Hamiltonian.
In particular, it is found that the localization effects due to repeated measurements are stronger than 
those under random scattering in the unitary evolution.
This approach offers a concept for the design, for instance, of quantum circuits. 
It can be extended by introducing a random basis, consisting of a mixture of localized and delocalized
eigenvectors or random-time measurements~\cite{Ziegler_2021,PhysRevA.103.022222}.


\end{document}